\documentclass[a4paper,11pt]{article}
\usepackage{pos}

\usepackage[symbol]{footmisc}
\usepackage[english]{babel}
\usepackage{graphicx}
\usepackage{graphics}
\usepackage{braket}
\usepackage{bbold}
\usepackage{amsmath}
\usepackage{nicefrac}
\usepackage{dcolumn}
\usepackage{bm}
\usepackage{slashed}
\usepackage{datetime}
\usepackage{mciteplus}
\usepackage{multirow}
\usepackage{bbold}
\usepackage{siunitx}
\usepackage{booktabs}
\usepackage{color, soul}
\usepackage[usenames,dvipsnames]{xcolor}
\usepackage{float}
\usepackage[utf8]{inputenc}
\usepackage[normalem]{ulem}
\usepackage{mathtools}
\usepackage{setspace}

\renewcommand{\thefootnote}{\fnsymbol{footnote}}

\newcommand{\LQCD}{\Lambda_{\rm QCD}}
\newcommand{\MSb}{\overline{\rm MS}}

\newcommand{\NLL}{{\rm NLL/NLO}}

\newcommand{\NLLs}{{\rm NLL/NLO^*}}
\newcommand{\NLLm}{{\rm NLL/NLO^-}}

\newcommand{\DY}{\Delta Y}

\title{NLL/NLO$^-$ studies on Higgs-plus-jet production with POWHEG+JETHAD}
\ShortTitle{NLL/NLO$^-$ Higgs-plus-jet with POWHEG+JETHAD}

\author*[a]{Francesco Giovanni Celiberto}
\author[b,c]{Luigi Delle Rose}
\author[d]{Michael Fucilla}
\author[b,c]{Gabriele Gatto}
\author[b,c]{Alessandro Papa}

\affiliation[a]{Universidad de Alcalá (UAH), Departamento de Física y Matemáticas, Campus Universitario, Alcalá de Henares, E-28805, Madrid, Spain}

\affiliation[b]{Dipartimento di Fisica, Università della Calabria, Arcavacata di Rende, I-87036, Cosenza, Italy}

\affiliation[c]{
INFN, Gruppo Collegato di Cosenza, Arcavacata di Rende, I-87036, Cosenza, Italy}

\affiliation[d]{Université Paris-Saclay, CNRS/IN2P3, IJCLab, 91405, Orsay, France}

\emailAdd{francesco.celiberto@uah.es}
\emailAdd{luigi.dellerose@unical.it}
\emailAdd{michael.fucilla@ijclab.in2p3.fr}
\emailAdd{gabriele.gatto@unical.it}
\emailAdd{alessandro.papa@fis.unical.it}

\abstract{We consider the semi-inclusive emission of a Higgs boson in association with a light-flavored jet separated by a large rapidity interval at the LHC.
The accessed kinematic regimes fall into the so-called semi-hard sector, whose theoretical description lies at the intersection corner between the collinear factorization and the high-energy resummation.
We present a prototype version of a matching procedure aimed at combining next-to-leading fixed-order (NLO) calculations from POWHEG with the resummation of next-to-leading energy logarithms (NLL) as obtained from JETHAD.}

\FullConference{16th International Symposium on Radiative Corrections: Applications of Quantum Field Theory to Phenomenology (RADCOR2023)\\
28th May - 2nd June, 2023\\
Crieff, Scotland, UK\\}

\begin{document}
\maketitle

\section{Introductory remarks}
\label{sec:introduction}

With the discovery of the Higgs boson at the LHC a new era of precision tests of the Standard Model, as well as of intensive searches for clues of New Physics, began.
In this respect, an accurate description of the gluon-gluon fusion channel in perturbative Quantum Chromodynamics (QCD)  is of top priority~\cite{Dawson:1990zj,Djouadi:1991tka}.
Higher-order calculations are necessary ingredients for precise studies of Higgs production \emph{via} the well-grounded \emph{collinear factorization}.
Here, cross sections are elegantly cast as one-dimensional convolutions between collinear parton distribution functions (PDFs) and on-shell perturbative coefficient functions.
At the same time, the theoretical description of Higgs-sensitive final states in the kinematic sectors accessible at the LHC and at future hadron and lepton colliders calls for the inclusion, to all orders, of logarithms which are systematically missed by a purely collinear vision. These logarithms can be large enough to spoil the convergence of the perturbative series, thus requiring the development of all-order \emph{resummation} techniques.

In this study we consider the \emph{semi-hard} QCD sector~\cite{Gribov:1983ivg,Celiberto:2017ius,Hentschinski:2022xnd}, where the rigorous scale hierarchy, $\sqrt{s} \gg \{Q\} \gg \LQCD $ ($\sqrt{s}$ is the squared center-of-mass energy, $\{Q\}$ is a set of process-dependent hard scales, $\LQCD$ is the QCD hadronization scale), brings to the growth of large energy logarithms.
The Balitsky--Fadin--Kuraev--Lipatov (BFKL) resummation~\cite{Fadin:1975cb,Balitsky:1978ic} offers us a systematic way to resum to all orders these logarithms within the leading-logarithmic (LL) and the next-to-leading logarithmic (NLL) level (for recent advancements beyond NLL, see Ref.~\cite{Caola:2021izf,Falcioni:2021dgr,Byrne:2022wzk,Fadin:2023roz}).
Remarkably, the BFKL formalism and its nonlinear extension to the saturation regime gives us a direct access to the gluon distribution in the nucleon at low-$x$~\cite{Bacchetta:2020vty,Arbuzov:2020cqg,Celiberto:2021zww,Celiberto:2022omz,Amoroso:2022eow,Bolognino:2018rhb,Bolognino:2021niq,Celiberto:2019slj,Peredo:2023oym,Taels:2022tza,Caucal:2023nci,Caucal:2023fsf}.
Suitable reactions whereby testing BFKL and, more in general, high-energy dynamics in hadron collisions, feature the semi-inclusive emission of two objects possessing high transverse masses and being strongly separated in rapidity.
One one hand, transverse masses well above $\LQCD$ make us fall into the semi-hard regime.
On the other hand, a large final-state rapidity interval, $\DY$, heightens the contribution of undetected gluons strongly ordered in rapidity, which are responsible for large logarithmic corrections.

A solid description of these two-particle hadroproduction channels calls for the employment of a \emph{multilateral} formalism, where both the collinear and the high-energy dynamics come into play. To this extent, a \emph{hybrid high-energy and collinear factorization} (HyF) was developed~\cite{Celiberto:2020tmb,Bolognino:2021mrc,Colferai:2023hje}.~\footnote{For similar approaches, close in spirit to ours, see Refs.~\cite{Deak:2009xt,vanHameren:2022mtk,Bonvini:2018ixe,Bonvini:2018iwt,Silvetti:2022hyc}.}
HyF partonic cross sections take the form of a convolution between two  impact factors (or emission functions), which are process-dependent, and the NLL BFKL Green's function (analogous to the Sudakov factor of soft-gluon resummations), which is the process-universal. 
Impact factors are in turn written as collinear convolutions between standard collinear PDFs and singly off-shell coefficient functions.
The state-of-the-art accuracy of HyF is NLL/NLO.
This means that, for a given process, the relevant coefficient functions need to be calculated at fixed NLO accuracy. 
Otherwise, one must rely upon a partial next-to-leading treatment, labeled as $\NLLs$ when only the Green's function is taken at NLL and both the coefficient functions are at LO, or $\NLLm$ when the Green's function is at NLL, one coefficient function is at NLO, and the other one is at LO.

Promising semi-inclusive channels whereby probing the semi-hard QCD sector are: emissions of two Mueller--Navelet jets~\cite{Ducloue:2013hia,Ducloue:2013bva,Caporale:2014gpa,Celiberto:2015yba,Celiberto:2015mpa,Celiberto:2016ygs,Caporale:2018qnm,Celiberto:2022gji}, multi-jet diffractive systems~\cite{Caporale:2016soq,Caporale:2016zkc,Caporale:2015int,Caporale:2016xku,Celiberto:2016vhn}, Drell--Yan pairs~\cite{Brzeminski:2016lwh,Celiberto:2018muu,Golec-Biernat:2018kem,Taels:2023czt}, light~\cite{Celiberto:2016hae,Celiberto:2017ptm,Celiberto:2016zgb,Bolognino:2018oth,Bolognino:2019yqj,Bolognino:2019cac,Celiberto:2020rxb,Celiberto:2022kxx} as well as singly heavy flavored~\cite{Celiberto:2017nyx,Bolognino:2019yls,Bolognino:2019ccd,AlexanderAryshev:2022pkx,Celiberto:2021dzy,Celiberto:2021fdp,Celiberto:2022zdg,Celiberto:2022keu,Anchordoqui:2021ghd,Feng:2022inv} hadrons, quarkonium states~\cite{Boussarie:2017oae,Chapon:2020heu,Celiberto:2022dyf,Celiberto:2023fzz,Stebel:2021bbn}, and exotic matter candidates~\cite{Celiberto:2023rzw}.
In this article we consider the semi-inclusive Higgs-plus-jet process, which was studied in perturbative QCD within next-to-NLO accuracy~\cite{Boughezal:2013uia,Chen:2014gva,Boughezal:2015dra} and \emph{via} the transverse-momentum resummation at the next-to-NLL level~\cite{Monni:2019yyr}.
As $\DY$ grows, the impact of energy logarithms becomes larger and larger.
Thus, the high-energy resummation, as encoded in the HyF formalism, comes out as a valuable tool for a proper and consistent description of Higgs-plus-jet differential rates~\cite{Celiberto:2020tmb,Celiberto:2023rtu,DelDuca:1993ga}.

We present the POWHEG+JETHAD method, a prototype version of a novel \emph{matching} procedure aimed at combining, in the context of Higgs-plus-jet rapidity and transverse-momentum distributions, next-to-leading fixed-order results with the resummation of next-to-leading energy logarithms.
Results presented in the next section are for Higgs-plus-jet rapidity and transverse momentum spectra with the matching accuracy pushed to $\NLLm$ accuracy. They supersede the $\NLLs$ predictions of Ref.~\cite{Celiberto:2023uuk_article}, but they are still preliminary, with a full $\NLL$ treatment being in preparation.

\section{Higgs-plus-jet production: Matching NLL to NLO}
\label{sec:matching}

\begin{figure*}[!t]
\centering

\includegraphics[scale=0.36,clip]{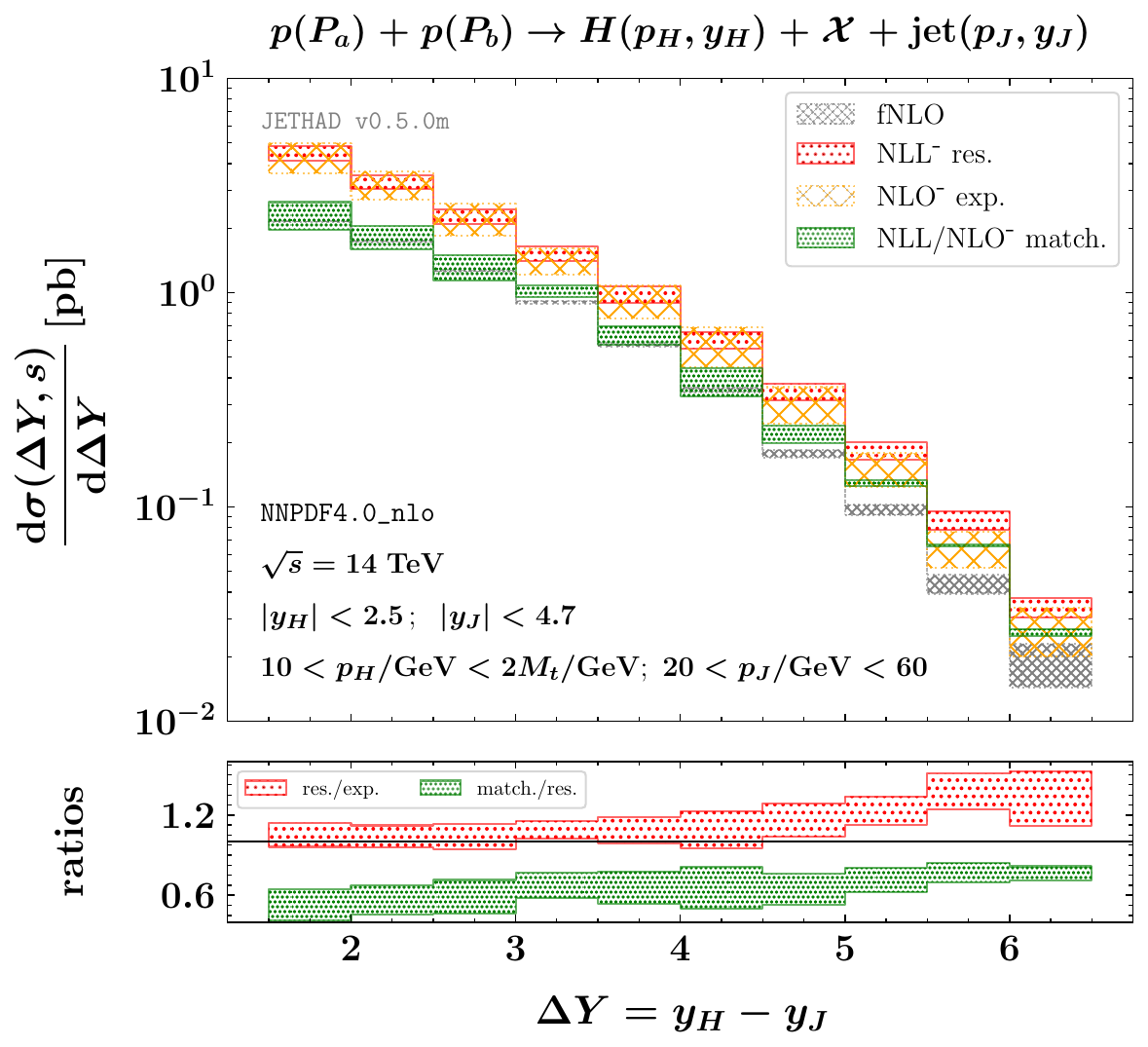}
 \hspace{0.30cm}
\includegraphics[scale=0.36,clip]{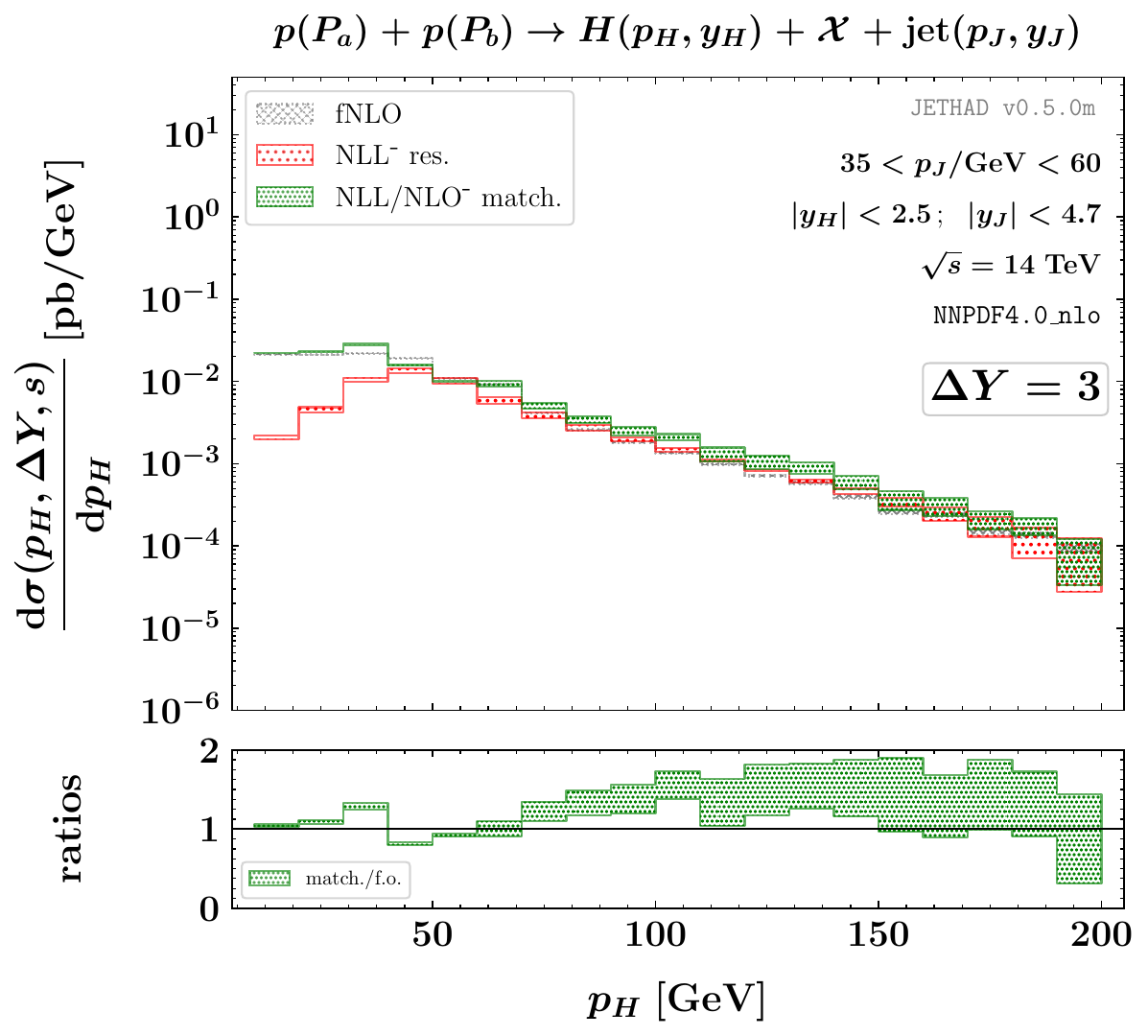}

\caption{Higgs-plus-jet rapidity (left) and transverse-momentum (right) rates at $\sqrt{s}=14\mbox{ TeV}$.
Uncertainty bands reflect the variation of $\mu_R$ and $\mu_F$ scales in the $1 < C_{\mu} < 2$ range. Text boxes exhibit kinematic cuts.
}

\label{fig:spectrum}
\end{figure*}

\begin{figure*}[!b]
\centering

\includegraphics[scale=0.36,clip]{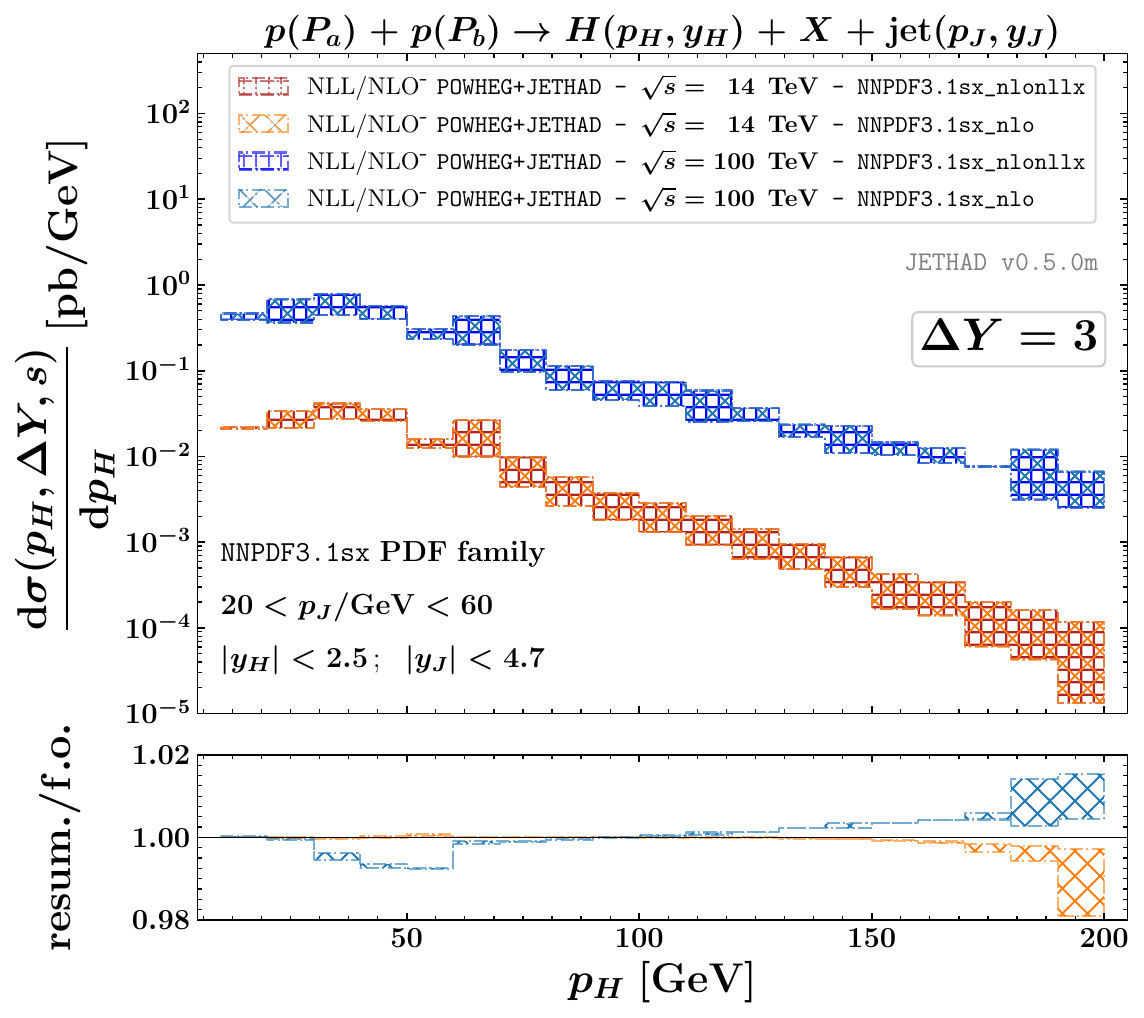}
 \hspace{0.60cm}
\includegraphics[scale=0.36,clip]{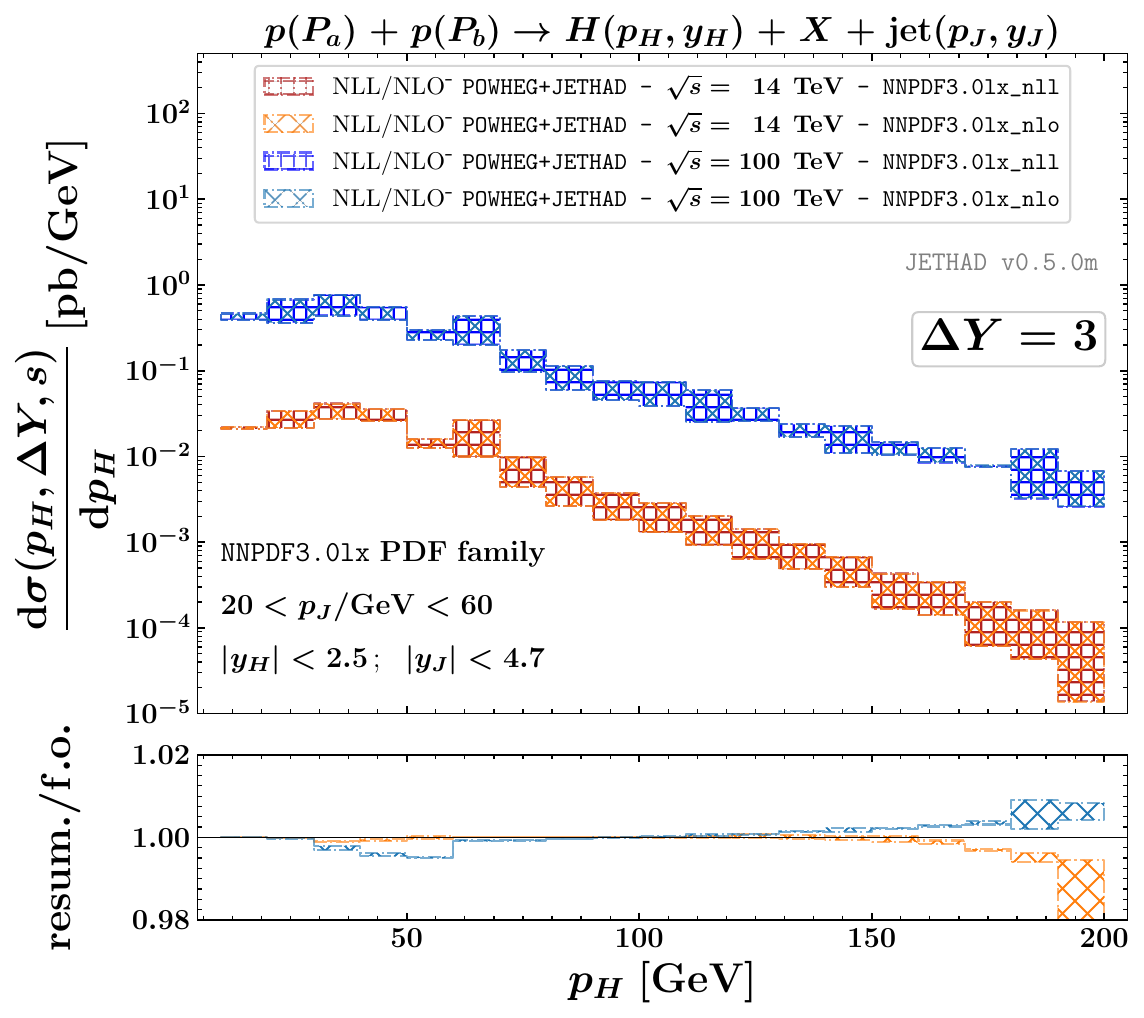}

\caption{Higgs-plus-jet transverse-momentum rate at $\sqrt{s}=14\mbox{ TeV}$. Left (right) plot shows the impact of a resummation-based low-$x$ (large-$x$) improvement on PDFs.
Uncertainty bands reflect the variation of $\mu_R$ and $\mu_F$ scales in the $1 < C_{\mu} < 2$ range. Text boxes exhibit kinematic cuts.
}

\label{fig:pH_spectrum_PDFs}
\end{figure*}

An insightful information coming from quite recent, HyF-related studies on the Higgs transverse-momentum ($p_H$) spectrum in semi-inclusive Higgs-plus-jet emissions at the LHC, is the solid stability which this distribution exhibits under higher-order corrections and energy-scale variations.
At the same time, however, large deviations HyF predictions from the fixed-order background have been observed, their weight reaching roughly two orders of magnitude when $p_H \gtrsim 120$~GeV~\cite{Celiberto:2020tmb}.
A similar trend has been shown by $\DY$-distributions at LHC as well as nominal FCC energies~\cite{Celiberto:2023rtu}.

This motivated us to develop a pioneering \emph{matching} procedure between NLO fixed-order results and NLL-resummed calculations, which permits to exactly remove, within the $\NLLm$ accuracy, the corresponding \emph{double counting}.
Indeed, given that the full NLO contribution to the forward Higgs emission function was calculated only recently~\cite{Celiberto:2022fgx,Fucilla:2023pma,Fucilla:2022whr,Hentschinski:2020tbi} and it has not yet been implemented in our reference technology, in the JETHAD code~\cite{Celiberto:2020wpk,Celiberto:2022rfj,Celiberto:2023fzz}, we will rely upon a $\NLLm$ treatment.
A sketch of our matching procedure reads

\begin{equation}
\label{eq:matching}
\begin{split}
 \hspace{-0.155cm}
 \underbrace{{\rm d}\sigma^{{{\rm NLL/NLO}}^{\boldsymbol{-}}}(\Delta Y, \varphi, s)}_{\text{\colorbox{OliveGreen}{\textbf{\textcolor{white}{NLL/NLO$^{\boldsymbol{-}}$}}} {\tt POWHEG+JETHAD}}} 
 = 
 \underbrace{{\rm d}\sigma^{\rm NLO}(\Delta Y, \varphi, s)}_{\text{\colorbox{gray}{\textcolor{white}{\textbf{NLO}}} {\tt POWHEG} w/o PS}}
 +\; 
 \underbrace{\underbrace{{\rm d}\sigma^{{{\rm NLL}}^{\boldsymbol{-}}}(\Delta Y, \varphi, s)}_{\text{\colorbox{red}{\textbf{\textcolor{white}{NLL$^{\boldsymbol{-}}$ resum}}} (HyF)}}
 \;-\; 
 \underbrace{\Delta{\rm d}\sigma^{{{\rm NLL/NLO}}^{\boldsymbol{-}}}(\Delta Y, \varphi, s)}_{\text{\colorbox{orange}{\textbf{NLL$^{\boldsymbol{-}}$ expanded}} at NLO}}}_{\text{\colorbox{NavyBlue}{\textbf{\textcolor{white}{NLL$^{\boldsymbol{-}}$}}} {\tt JETHAD} w/o NLO$^{\boldsymbol{-}}$ double counting}}
 \,.
\end{split}
\end{equation}

A given differential cross section, matched at $\NLLm$ (green) \emph{via} the POWHEG+JETHAD method, takes the form of a sum of the NLO fixed-order contribution (gray) as obtained from the POWHEG technology~\cite{Nason:2004rx,Campbell:2012am,Hamilton:2012rf,Banfi:2023mhz,Bagnaschi:2023rbx} and the $\rm NLL^-$ resummed part (blue) from JETHAD. The latter is given by the $\rm NLL^-$ HyF resummed contribution (red) minus the $\rm NLL^-$ expanded (orange) at NLO, \emph{i.e.} without the doubly-counted term.
Removing it from inside JETHAD instead of POWHEG makes our procedure dynamically compatible with other possible matching formalism. More importantly, it allows us to discard spurious power-correction contaminations genuinely accounted for by HyF to all orders.
We remark that POWHEG has been employed to calculate the fixed-order background, \emph{i.e.} without adding \emph{parton-shower} (PS) effects~\cite{Alioli:2022dkj,Alioli:2023har,Buckley:2021gfw,vanBeekveld:2022zhl,vanBeekveld:2022ukn,FerrarioRavasio:2023kyg}.

Figure~\ref{fig:spectrum} contains preliminary $\NLLm$ results for the $\DY$ (left) and $p_H$ (right) spectra at 14~TeV LHC.
Calculations were performed in the $\MSb$ scheme, and {\tt NNPDF4.0\_nlo} collinear PDFs were adopted~\cite{NNPDF:2021uiq,NNPDF:2021njg}.
The color code in Fig.~\ref{fig:spectrum} matches the one of Eq.~\eqref{eq:matching}. Ancillary panels below primary plots show the reliability of our matching. In particular, focusing on the $\DY$ spectrum (left), the NLL-resummed contribution is very small when compared with the expanded term at low $\DY$, while their ratio (red) generally increases with $\DY$. Furthermore, the matched-over-resummed ratio (green) is smaller than one at low $\DY$, and tends to one in the large $\DY$ range.
All this clearly indicates that our matching is catching the core dynamics of our process, with the high-energy resummation becoming more and more relevant as $\DY$ increases, as expected.

Figure~\ref{fig:pH_spectrum_PDFs} provides us with an additional analysis on the $p_H$ spectrum, at $\DY=3$, and with small-$x$ (left) or large-$x$ (right) resummation improvements on collinear PDFs at 14~TeV~LHC and 100~TeV~FCC energies.
Left panel is for $p_H$ distributions obtained by making use of small-$x$ resummed PDFs from the {\tt NNPDF3.1sx} family~\cite{Ball:2017otu}, whereas right panel shows transverse-momentum rates obtained by means of large-$x$, threshold resummed PDFs from the {\tt NNPDF3.0lx} one~\cite{Bonvini:2015ira}.
Ancillary panels below primary plots clearly indicate that the overall effects is relatively small, globally staying below 2\%.
For both resummations they are more pronounced and negative in the peak region, $30 \lesssim p_H/{\rm GeV} \lesssim 60$, but only in the FCC case (turquoise), while they change sign in the large-$p_H$ tail, being negative at LHC energies and then becoming positive at FCC ones.
We stress that our study on the large-$x$ improvement should be intended as a proxy for the effect of the threshold resummation~\cite{Sterman:1986aj,Catani:1989ne,Catani:1996yz,Bonciani:2003nt,deFlorian:2012yg,Forte:2021wxe,Mukherjee:2006uu,Becher:2006nr,Becher:2007ty,Bonvini:2010tp,Ahmed:2014era,Banerjee:2018vvb,Ajjath:2021bbm,Duhr:2022cob} coming from PDFs only. To quantify the full impact of the threshold resummation on our high-energy observables, know to be sizable~\cite{Bolognino:2018oth,Celiberto:2020rxb,Celiberto:2020wpk}, one must develop a systematic method to resum large-$x$ logarithms in our off-shell coefficient functions.

\section{Conclusions and Outlook}
\label{sec:conclusions}

We developed a prototype version of a matching procedure, relying on the POWHEG~\cite{Nason:2004rx,Campbell:2012am,Hamilton:2012rf,Banfi:2023mhz,Bagnaschi:2023rbx} and JETHAD~\cite{Celiberto:2020wpk,Celiberto:2022rfj,Celiberto:2023fzz} codes. It purpose is combining NLO fixed-order calculations with the high-energy resummation at NLL.
Future works will extend this study to: $a)$ gauge the size of full NLO contributions~\cite{Celiberto:2022fgx,Hentschinski:2020tbi}, $b)$ assess the weight of heavy-quark finite-mass corrections~\cite{Jones:2018hbb,Bonciani:2022jmb}, $c)$ compare our predictions with PS~\cite{Alioli:2023har,Alioli:2022dkj,Buckley:2021gfw,vanBeekveld:2022zhl,vanBeekveld:2022ukn,FerrarioRavasio:2023kyg} and HEJ~\cite{Andersen:2022zte,Andersen:2023kuj} inspired ones.

\begingroup
\setstretch{0.6}
\bibliographystyle{bibstyle}
\bibliography{references}

\begin{thebibliography}{118}
\expandafter\ifx\csname natexlab\endcsname\relax\def\natexlab#1{#1}\fi
\expandafter\ifx\csname bibnamefont\endcsname\relax
  \def\bibnamefont#1{#1}\fi
\expandafter\ifx\csname bibfnamefont\endcsname\relax
  \def\bibfnamefont#1{#1}\fi
\expandafter\ifx\csname citenamefont\endcsname\relax
  \def\citenamefont#1{#1}\fi
\expandafter\ifx\csname url\endcsname\relax
  \def\url#1{\texttt{#1}}\fi
\expandafter\ifx\csname urlprefix\endcsname\relax\def\urlprefix{URL }\fi
\providecommand{\bibinfo}[2]{#2}
\providecommand{\eprint}[2][]{\url{#2}}

\bibitem[{\citenamefont{Dawson}(1991)}]{Dawson:1990zj}
\bibinfo{author}{\bibfnamefont{S.}~\bibnamefont{Dawson}}, \bibinfo{journal}{Nucl. Phys. B} \textbf{\bibinfo{volume}{359}}, \bibinfo{pages}{283} (\bibinfo{year}{1991}).

\bibitem[{\citenamefont{Djouadi et~al.}(1991)\citenamefont{Djouadi, Spira,  Zerwas}}]{Djouadi:1991tka}
\bibinfo{author}{\bibfnamefont{A.}~\bibnamefont{Djouadi}}, \bibinfo{author}{\bibfnamefont{M.}~\bibnamefont{Spira}},  \bibinfo{author}{\bibfnamefont{P.~M.} \bibnamefont{Zerwas}}, \bibinfo{journal}{Phys. Lett. B} \textbf{\bibinfo{volume}{264}}, \bibinfo{pages}{440} (\bibinfo{year}{1991}).

\bibitem[{\citenamefont{Gribov et~al.}(1983)\citenamefont{Gribov, Levin,  Ryskin}}]{Gribov:1983ivg}
\bibinfo{author}{\bibfnamefont{L.~V.} \bibnamefont{Gribov}}, \bibinfo{author}{\bibfnamefont{E.~M.} \bibnamefont{Levin}},  \bibinfo{author}{\bibfnamefont{M.~G.} \bibnamefont{Ryskin}}, \bibinfo{journal}{Phys. Rept.} \textbf{\bibinfo{volume}{100}}, \bibinfo{pages}{1} (\bibinfo{year}{1983}).

\bibitem[{\citenamefont{Celiberto}(2017)}]{Celiberto:2017ius}
\bibinfo{author}{\bibfnamefont{F.~G.} \bibnamefont{Celiberto}}, Ph.D. thesis (\bibinfo{year}{2017}), \eprint{1707.04315}.

\bibitem[{\citenamefont{Hentschinski et~al.}(2023)}]{Hentschinski:2022xnd}
\bibinfo{author}{\bibfnamefont{M.}~\bibnamefont{Hentschinski}} \bibnamefont{et~al.}, \bibinfo{journal}{Acta Phys. Polon. B} \textbf{\bibinfo{volume}{54}}, \bibinfo{pages}{2} (\bibinfo{year}{2023}), \eprint{2203.08129}.

\bibitem[{\citenamefont{Fadin et~al.}(1975)}]{Fadin:1975cb}
\bibinfo{author}{\bibfnamefont{V.~S.} \bibnamefont{Fadin}} \bibnamefont{et~al.}, \bibinfo{journal}{Phys. Lett. B} \textbf{\bibinfo{volume}{60}}, \bibinfo{pages}{50} (\bibinfo{year}{1975}).

\bibitem[{\citenamefont{Balitsky  Lipatov}(1978)}]{Balitsky:1978ic}
\bibinfo{author}{\bibfnamefont{I.}~\bibnamefont{Balitsky}}  \bibinfo{author}{\bibfnamefont{L.}~\bibnamefont{Lipatov}}, \bibinfo{journal}{Sov.\ J.\ Nucl.\ Phys.} \textbf{\bibinfo{volume}{28}}, \bibinfo{pages}{822} (\bibinfo{year}{1978}).

\bibitem[{\citenamefont{Caola et~al.}(2022)}]{Caola:2021izf}
\bibinfo{author}{\bibfnamefont{F.}~\bibnamefont{Caola}} \bibnamefont{et~al.}, \bibinfo{journal}{Phys. Rev. Lett.} \textbf{\bibinfo{volume}{128}}, \bibinfo{pages}{212001} (\bibinfo{year}{2022}), \eprint{2112.11097}.

\bibitem[{\citenamefont{Falcioni et~al.}(2022)}]{Falcioni:2021dgr}
\bibinfo{author}{\bibfnamefont{G.}~\bibnamefont{Falcioni}} \bibnamefont{et~al.}, \bibinfo{journal}{Phys. Rev. Lett.} \textbf{\bibinfo{volume}{128}}, \bibinfo{pages}{132001} (\bibinfo{year}{2022}), \eprint{2112.11098}.

\bibitem[{\citenamefont{Byrne et~al.}(2022)}]{Byrne:2022wzk}
\bibinfo{author}{\bibfnamefont{E.~P.} \bibnamefont{Byrne}} \bibnamefont{et~al.}, \bibinfo{journal}{JHEP} \textbf{\bibinfo{volume}{08}}, \bibinfo{pages}{271} (\bibinfo{year}{2022}), \eprint{2204.12459}.

\bibitem[{\citenamefont{Fadin et~al.}(2023)}]{Fadin:2023roz}
\bibinfo{author}{\bibfnamefont{V.~S.} \bibnamefont{Fadin}} \bibnamefont{et~al.}, \bibinfo{journal}{JHEP} \textbf{\bibinfo{volume}{04}}, \bibinfo{pages}{137} (\bibinfo{year}{2023}), \eprint{2302.09868}.

\bibitem[{\citenamefont{Bacchetta et~al.}(2020)}]{Bacchetta:2020vty}
\bibinfo{author}{\bibfnamefont{A.}~\bibnamefont{Bacchetta}} \bibnamefont{et~al.}, \bibinfo{journal}{Eur. Phys. J. C} \textbf{\bibinfo{volume}{80}}, \bibinfo{pages}{733} (\bibinfo{year}{2020}), \eprint{2005.02288}.

\bibitem[{\citenamefont{Arbuzov et~al.}(2021)}]{Arbuzov:2020cqg}
\bibinfo{author}{\bibfnamefont{A.}~\bibnamefont{Arbuzov}} \bibnamefont{et~al.}, \bibinfo{journal}{Prog. Part. Nucl. Phys.} \textbf{\bibinfo{volume}{119}}, \bibinfo{pages}{103858} (\bibinfo{year}{2021}), \eprint{2011.15005}.

\bibitem[{\citenamefont{Celiberto}(2021{\natexlab{a}})}]{Celiberto:2021zww}
\bibinfo{author}{\bibfnamefont{F.~G.} \bibnamefont{Celiberto}}, \bibinfo{journal}{Nuovo Cim.} \textbf{\bibinfo{volume}{C44}}, \bibinfo{pages}{36} (\bibinfo{year}{2021}{\natexlab{a}}), \eprint{2101.04630}.

\bibitem[{\citenamefont{Celiberto}(2022{\natexlab{a}})}]{Celiberto:2022omz}
\bibinfo{author}{\bibfnamefont{F.~G.} \bibnamefont{Celiberto}}, \bibinfo{journal}{Universe} \textbf{\bibinfo{volume}{8}}, \bibinfo{pages}{661} (\bibinfo{year}{2022}{\natexlab{a}}), \eprint{2210.08322}.

\bibitem[{\citenamefont{Amoroso et~al.}(2022)}]{Amoroso:2022eow}
\bibinfo{author}{\bibfnamefont{S.}~\bibnamefont{Amoroso}} \bibnamefont{et~al.}, \bibinfo{journal}{Acta Phys. Polon. B} \textbf{\bibinfo{volume}{53}}, \bibinfo{pages}{A1} (\bibinfo{year}{2022}), \eprint{2203.13923}.

\bibitem[{\citenamefont{Bolognino et~al.}(2018{\natexlab{a}})}]{Bolognino:2018rhb}
\bibinfo{author}{\bibfnamefont{A.~D.} \bibnamefont{Bolognino}} \bibnamefont{et~al.}, \bibinfo{journal}{Eur. Phys. J.} \textbf{\bibinfo{volume}{C78}}, \bibinfo{pages}{1023} (\bibinfo{year}{2018}{\natexlab{a}}), \eprint{1808.02395}.

\bibitem[{\citenamefont{Bolognino et~al.}(2021{\natexlab{a}})}]{Bolognino:2021niq}
\bibinfo{author}{\bibfnamefont{A.~D.} \bibnamefont{Bolognino}} \bibnamefont{et~al.}, \bibinfo{journal}{Eur. Phys. J. C} \textbf{\bibinfo{volume}{81}}, \bibinfo{pages}{846} (\bibinfo{year}{2021}{\natexlab{a}}), \eprint{2107.13415}.

\bibitem[{\citenamefont{Celiberto}(2019)}]{Celiberto:2019slj}
\bibinfo{author}{\bibfnamefont{F.~G.} \bibnamefont{Celiberto}}, \bibinfo{journal}{Nuovo Cim.} \textbf{\bibinfo{volume}{C42}}, \bibinfo{pages}{220} (\bibinfo{year}{2019}), \eprint{1912.11313}.

\bibitem[{\citenamefont{Peredo  Hentschinski}(2023)}]{Peredo:2023oym}
\bibinfo{author}{\bibfnamefont{M.~A.} \bibnamefont{Peredo}}  \bibinfo{author}{\bibfnamefont{M.}~\bibnamefont{Hentschinski}} (\bibinfo{year}{2023}), \eprint{2308.15430}.

\bibitem[{\citenamefont{Taels et~al.}(2022)}]{Taels:2022tza}
\bibinfo{author}{\bibfnamefont{P.}~\bibnamefont{Taels}} \bibnamefont{et~al.}, \bibinfo{journal}{JHEP} \textbf{\bibinfo{volume}{10}}, \bibinfo{pages}{184} (\bibinfo{year}{2022}), \eprint{2204.11650}.

\bibitem[{\citenamefont{Caucal et~al.}(2023{\natexlab{a}})}]{Caucal:2023nci}
\bibinfo{author}{\bibfnamefont{P.}~\bibnamefont{Caucal}} \bibnamefont{et~al.}, \bibinfo{journal}{JHEP} \textbf{\bibinfo{volume}{08}}, \bibinfo{pages}{062} (\bibinfo{year}{2023}{\natexlab{a}}), \eprint{2304.03304}.

\bibitem[{\citenamefont{Caucal et~al.}(2023{\natexlab{b}})}]{Caucal:2023fsf}
\bibinfo{author}{\bibfnamefont{P.}~\bibnamefont{Caucal}} \bibnamefont{et~al.} (\bibinfo{year}{2023}{\natexlab{b}}), \eprint{2308.00022}.

\bibitem[{\citenamefont{Celiberto et~al.}(2021{\natexlab{a}})}]{Celiberto:2020tmb}
\bibinfo{author}{\bibfnamefont{F.~G.} \bibnamefont{Celiberto}} \bibnamefont{et~al.}, \bibinfo{journal}{Eur. Phys. J. C} \textbf{\bibinfo{volume}{81}}, \bibinfo{pages}{293} (\bibinfo{year}{2021}{\natexlab{a}}), \eprint{2008.00501}.

\bibitem[{\citenamefont{Bolognino et~al.}(2021{\natexlab{b}})}]{Bolognino:2021mrc}
\bibinfo{author}{\bibfnamefont{A.~D.} \bibnamefont{Bolognino}} \bibnamefont{et~al.}, \bibinfo{journal}{Phys. Rev. D} \textbf{\bibinfo{volume}{103}}, \bibinfo{pages}{094004} (\bibinfo{year}{2021}{\natexlab{b}}), \eprint{2103.07396}.

\bibitem[{\citenamefont{Colferai et~al.}(2023)}]{Colferai:2023hje}
\bibinfo{author}{\bibfnamefont{D.}~\bibnamefont{Colferai}} \bibnamefont{et~al.}, \bibinfo{journal}{JHEP} \textbf{\bibinfo{volume}{06}}, \bibinfo{pages}{091} (\bibinfo{year}{2023}), \eprint{2304.09073}.

\bibitem[{\citenamefont{Deak et~al.}(2009)}]{Deak:2009xt}
\bibinfo{author}{\bibfnamefont{M.}~\bibnamefont{Deak}} \bibnamefont{et~al.}, \bibinfo{journal}{JHEP} \textbf{\bibinfo{volume}{09}}, \bibinfo{pages}{121} (\bibinfo{year}{2009}), \eprint{0908.0538}.

\bibitem[{\citenamefont{van Hameren et~al.}(2022)\citenamefont{van Hameren, Motyka,  Ziarko}}]{vanHameren:2022mtk}
\bibinfo{author}{\bibfnamefont{A.}~\bibnamefont{van Hameren}}, \bibinfo{author}{\bibfnamefont{L.}~\bibnamefont{Motyka}},  \bibinfo{author}{\bibfnamefont{G.}~\bibnamefont{Ziarko}}, \bibinfo{journal}{JHEP} \textbf{\bibinfo{volume}{11}}, \bibinfo{pages}{103} (\bibinfo{year}{2022}), \eprint{2205.09585}.

\bibitem[{\citenamefont{Bonvini  Marzani}(2018)}]{Bonvini:2018ixe}
\bibinfo{author}{\bibfnamefont{M.}~\bibnamefont{Bonvini}}  \bibinfo{author}{\bibfnamefont{S.}~\bibnamefont{Marzani}}, \bibinfo{journal}{Phys. Rev. Lett.} \textbf{\bibinfo{volume}{120}}, \bibinfo{pages}{202003} (\bibinfo{year}{2018}), \eprint{1802.07758}.

\bibitem[{\citenamefont{Bonvini}(2018)}]{Bonvini:2018iwt}
\bibinfo{author}{\bibfnamefont{M.}~\bibnamefont{Bonvini}}, \bibinfo{journal}{Eur. Phys. J. C} \textbf{\bibinfo{volume}{78}}, \bibinfo{pages}{834} (\bibinfo{year}{2018}), \eprint{1805.08785}.

\bibitem[{\citenamefont{Silvetti  Bonvini}(2023)}]{Silvetti:2022hyc}
\bibinfo{author}{\bibfnamefont{F.}~\bibnamefont{Silvetti}}  \bibinfo{author}{\bibfnamefont{M.}~\bibnamefont{Bonvini}}, \bibinfo{journal}{Eur. Phys. J. C} \textbf{\bibinfo{volume}{83}}, \bibinfo{pages}{267} (\bibinfo{year}{2023}), \eprint{2211.10142}.

\bibitem[{\citenamefont{Duclou\'e et~al.}(2013)\citenamefont{Duclou\'e, Szymanowski,  Wallon}}]{Ducloue:2013hia}
\bibinfo{author}{\bibfnamefont{B.}~\bibnamefont{Duclou\'e}}, \bibinfo{author}{\bibfnamefont{L.}~\bibnamefont{Szymanowski}},  \bibinfo{author}{\bibfnamefont{S.}~\bibnamefont{Wallon}}, \bibinfo{journal}{JHEP} \textbf{\bibinfo{volume}{05}}, \bibinfo{pages}{096} (\bibinfo{year}{2013}), \eprint{1302.7012}.

\bibitem[{\citenamefont{Duclou\'e et~al.}(2014)\citenamefont{Duclou\'e, Szymanowski,  Wallon}}]{Ducloue:2013bva}
\bibinfo{author}{\bibfnamefont{B.}~\bibnamefont{Duclou\'e}}, \bibinfo{author}{\bibfnamefont{L.}~\bibnamefont{Szymanowski}},  \bibinfo{author}{\bibfnamefont{S.}~\bibnamefont{Wallon}}, \bibinfo{journal}{Phys. Rev. Lett.} \textbf{\bibinfo{volume}{112}}, \bibinfo{pages}{082003} (\bibinfo{year}{2014}), \eprint{1309.3229}.

\bibitem[{\citenamefont{Caporale et~al.}(2014)}]{Caporale:2014gpa}
\bibinfo{author}{\bibfnamefont{F.}~\bibnamefont{Caporale}} \bibnamefont{et~al.}, \bibinfo{journal}{Eur. Phys. J. C} \textbf{\bibinfo{volume}{74}}, \bibinfo{pages}{3084} (\bibinfo{year}{2014}), \eprint{1407.8431}.

\bibitem[{\citenamefont{Celiberto et~al.}(2015{\natexlab{a}})}]{Celiberto:2015yba}
\bibinfo{author}{\bibfnamefont{F.~G.} \bibnamefont{Celiberto}} \bibnamefont{et~al.}, \bibinfo{journal}{Eur. Phys. J. C} \textbf{\bibinfo{volume}{75}}, \bibinfo{pages}{292} (\bibinfo{year}{2015}{\natexlab{a}}), \eprint{1504.08233}.

\bibitem[{\citenamefont{Celiberto et~al.}(2015{\natexlab{b}})}]{Celiberto:2015mpa}
\bibinfo{author}{\bibfnamefont{F.~G.} \bibnamefont{Celiberto}} \bibnamefont{et~al.}, \bibinfo{journal}{Acta Phys. Polon. Supp.} \textbf{\bibinfo{volume}{8}}, \bibinfo{pages}{935} (\bibinfo{year}{2015}{\natexlab{b}}), \eprint{1510.01626}.

\bibitem[{\citenamefont{Celiberto et~al.}(2016{\natexlab{a}})}]{Celiberto:2016ygs}
\bibinfo{author}{\bibfnamefont{F.~G.} \bibnamefont{Celiberto}} \bibnamefont{et~al.}, \bibinfo{journal}{Eur. Phys. J. C} \textbf{\bibinfo{volume}{76}}, \bibinfo{pages}{224} (\bibinfo{year}{2016}{\natexlab{a}}), \eprint{1601.07847}.

\bibitem[{\citenamefont{Caporale et~al.}(2018)}]{Caporale:2018qnm}
\bibinfo{author}{\bibfnamefont{F.}~\bibnamefont{Caporale}} \bibnamefont{et~al.}, \bibinfo{journal}{Nucl. Phys. B} \textbf{\bibinfo{volume}{935}}, \bibinfo{pages}{412} (\bibinfo{year}{2018}), \eprint{1806.06309}.

\bibitem[{\citenamefont{Celiberto et~al.}(2022{\natexlab{a}})}]{Celiberto:2022gji}
\bibinfo{author}{\bibfnamefont{F.~G.} \bibnamefont{Celiberto}} \bibnamefont{et~al.}, \bibinfo{journal}{Phys. Rev. D} \textbf{\bibinfo{volume}{106}}, \bibinfo{pages}{114004} (\bibinfo{year}{2022}{\natexlab{a}}), \eprint{2207.05015}.

\bibitem[{\citenamefont{Caporale et~al.}(2016{\natexlab{a}})}]{Caporale:2016soq}
\bibinfo{author}{\bibfnamefont{F.}~\bibnamefont{Caporale}} \bibnamefont{et~al.}, \bibinfo{journal}{Nucl. Phys. B} \textbf{\bibinfo{volume}{910}}, \bibinfo{pages}{374} (\bibinfo{year}{2016}{\natexlab{a}}), \eprint{1603.07785}.

\bibitem[{\citenamefont{Caporale et~al.}(2017{\natexlab{a}})}]{Caporale:2016zkc}
\bibinfo{author}{\bibfnamefont{F.}~\bibnamefont{Caporale}} \bibnamefont{et~al.}, \bibinfo{journal}{Phys. Rev. D} \textbf{\bibinfo{volume}{95}}, \bibinfo{pages}{074007} (\bibinfo{year}{2017}{\natexlab{a}}), \eprint{1612.05428}.

\bibitem[{\citenamefont{Caporale et~al.}(2016{\natexlab{b}})}]{Caporale:2015int}
\bibinfo{author}{\bibfnamefont{F.}~\bibnamefont{Caporale}} \bibnamefont{et~al.}, \bibinfo{journal}{Eur. Phys. J. C} \textbf{\bibinfo{volume}{76}}, \bibinfo{pages}{165} (\bibinfo{year}{2016}{\natexlab{b}}), \eprint{1512.03364}.

\bibitem[{\citenamefont{Caporale et~al.}(2017{\natexlab{b}})}]{Caporale:2016xku}
\bibinfo{author}{\bibfnamefont{F.}~\bibnamefont{Caporale}} \bibnamefont{et~al.}, \bibinfo{journal}{Eur. Phys. J. C} \textbf{\bibinfo{volume}{77}}, \bibinfo{pages}{5} (\bibinfo{year}{2017}{\natexlab{b}}), \eprint{1606.00574}.

\bibitem[{\citenamefont{Celiberto}(2016)}]{Celiberto:2016vhn}
\bibinfo{author}{\bibfnamefont{F.~G.} \bibnamefont{Celiberto}}, \bibinfo{journal}{Frascati Phys. Ser.} \textbf{\bibinfo{volume}{63}}, \bibinfo{pages}{43} (\bibinfo{year}{2016}), \eprint{1606.07327}.

\bibitem[{\citenamefont{Brzeminski et~al.}(2017)}]{Brzeminski:2016lwh}
\bibinfo{author}{\bibfnamefont{D.}~\bibnamefont{Brzeminski}} \bibnamefont{et~al.}, \bibinfo{journal}{JHEP} \textbf{\bibinfo{volume}{01}}, \bibinfo{pages}{005} (\bibinfo{year}{2017}), \eprint{1611.04449}.

\bibitem[{\citenamefont{Celiberto et~al.}(2018{\natexlab{a}})}]{Celiberto:2018muu}
\bibinfo{author}{\bibfnamefont{F.~G.} \bibnamefont{Celiberto}} \bibnamefont{et~al.}, \bibinfo{journal}{Phys. Lett.} \textbf{\bibinfo{volume}{B786}}, \bibinfo{pages}{201} (\bibinfo{year}{2018}{\natexlab{a}}), \eprint{1808.09511}.

\bibitem[{\citenamefont{Golec-Biernat et~al.}(2018)}]{Golec-Biernat:2018kem}
\bibinfo{author}{\bibfnamefont{K.}~\bibnamefont{Golec-Biernat}} \bibnamefont{et~al.}, \bibinfo{journal}{JHEP} \textbf{\bibinfo{volume}{12}}, \bibinfo{pages}{091} (\bibinfo{year}{2018}), \eprint{1811.04361}.

\bibitem[{\citenamefont{Taels}(2023)}]{Taels:2023czt}
\bibinfo{author}{\bibfnamefont{P.}~\bibnamefont{Taels}} (\bibinfo{year}{2023}), \eprint{2308.02449}.

\bibitem[{\citenamefont{Celiberto et~al.}(2016{\natexlab{b}})}]{Celiberto:2016hae}
\bibinfo{author}{\bibfnamefont{F.~G.} \bibnamefont{Celiberto}} \bibnamefont{et~al.}, \bibinfo{journal}{Phys. Rev. D} \textbf{\bibinfo{volume}{94}}, \bibinfo{pages}{034013} (\bibinfo{year}{2016}{\natexlab{b}}), \eprint{1604.08013}.

\bibitem[{\citenamefont{Celiberto et~al.}(2017{\natexlab{a}})}]{Celiberto:2017ptm}
\bibinfo{author}{\bibfnamefont{F.~G.} \bibnamefont{Celiberto}} \bibnamefont{et~al.}, \bibinfo{journal}{Eur. Phys. J. C} \textbf{\bibinfo{volume}{77}}, \bibinfo{pages}{382} (\bibinfo{year}{2017}{\natexlab{a}}), \eprint{1701.05077}.

\bibitem[{\citenamefont{Celiberto et~al.}(2017{\natexlab{b}})}]{Celiberto:2016zgb}
\bibinfo{author}{\bibfnamefont{F.~G.} \bibnamefont{Celiberto}} \bibnamefont{et~al.}, \bibinfo{journal}{AIP Conf. Proc.} \textbf{\bibinfo{volume}{1819}}, \bibinfo{pages}{060005} (\bibinfo{year}{2017}{\natexlab{b}}), \eprint{1611.04811}.

\bibitem[{\citenamefont{Bolognino et~al.}(2018{\natexlab{b}})}]{Bolognino:2018oth}
\bibinfo{author}{\bibfnamefont{A.~D.} \bibnamefont{Bolognino}} \bibnamefont{et~al.}, \bibinfo{journal}{Eur. Phys. J. C} \textbf{\bibinfo{volume}{78}}, \bibinfo{pages}{772} (\bibinfo{year}{2018}{\natexlab{b}}), \eprint{1808.05483}.

\bibitem[{\citenamefont{Bolognino et~al.}(2019{\natexlab{a}})}]{Bolognino:2019yqj}
\bibinfo{author}{\bibfnamefont{A.~D.} \bibnamefont{Bolognino}} \bibnamefont{et~al.}, \bibinfo{journal}{Acta Phys. Polon. Supp.} \textbf{\bibinfo{volume}{12}}, \bibinfo{pages}{773} (\bibinfo{year}{2019}{\natexlab{a}}), \eprint{1902.04511}.

\bibitem[{\citenamefont{Bolognino et~al.}(2019{\natexlab{b}})}]{Bolognino:2019cac}
\bibinfo{author}{\bibfnamefont{A.~D.} \bibnamefont{Bolognino}} \bibnamefont{et~al.}, \bibinfo{journal}{PoS} \textbf{\bibinfo{volume}{DIS2019}}, \bibinfo{pages}{049} (\bibinfo{year}{2019}{\natexlab{b}}), \eprint{1906.11800}.

\bibitem[{\citenamefont{Celiberto et~al.}(2020)\citenamefont{Celiberto, Ivanov,  Papa}}]{Celiberto:2020rxb}
\bibinfo{author}{\bibfnamefont{F.~G.} \bibnamefont{Celiberto}}, \bibinfo{author}{\bibfnamefont{D.~{\relax Yu}.} \bibnamefont{Ivanov}},  \bibinfo{author}{\bibfnamefont{A.}~\bibnamefont{Papa}}, \bibinfo{journal}{Phys. Rev. D} \textbf{\bibinfo{volume}{102}}, \bibinfo{pages}{094019} (\bibinfo{year}{2020}), \eprint{2008.10513}.

\bibitem[{\citenamefont{Celiberto}(2023{\natexlab{a}})}]{Celiberto:2022kxx}
\bibinfo{author}{\bibfnamefont{F.~G.} \bibnamefont{Celiberto}}, \bibinfo{journal}{Eur. Phys. J. C} \textbf{\bibinfo{volume}{83}}, \bibinfo{pages}{332} (\bibinfo{year}{2023}{\natexlab{a}}), \eprint{2208.14577}.

\bibitem[{\citenamefont{Celiberto et~al.}(2018{\natexlab{b}})}]{Celiberto:2017nyx}
\bibinfo{author}{\bibfnamefont{F.~G.} \bibnamefont{Celiberto}} \bibnamefont{et~al.}, \bibinfo{journal}{Phys. Lett. B} \textbf{\bibinfo{volume}{777}}, \bibinfo{pages}{141} (\bibinfo{year}{2018}{\natexlab{b}}), \eprint{1709.10032}.

\bibitem[{\citenamefont{Bolognino et~al.}(2019{\natexlab{c}})}]{Bolognino:2019yls}
\bibinfo{author}{\bibfnamefont{A.~D.} \bibnamefont{Bolognino}} \bibnamefont{et~al.}, \bibinfo{journal}{Eur. Phys. J. C} \textbf{\bibinfo{volume}{79}}, \bibinfo{pages}{939} (\bibinfo{year}{2019}{\natexlab{c}}), \eprint{1909.03068}.

\bibitem[{\citenamefont{Bolognino et~al.}(2019{\natexlab{d}})}]{Bolognino:2019ccd}
\bibinfo{author}{\bibfnamefont{A.~D.} \bibnamefont{Bolognino}} \bibnamefont{et~al.}, \bibinfo{journal}{PoS} \textbf{\bibinfo{volume}{DIS2019}}, \bibinfo{pages}{067} (\bibinfo{year}{2019}{\natexlab{d}}), \eprint{1906.05940}.

\bibitem[{\citenamefont{Adachi et~al.}(2022)}]{AlexanderAryshev:2022pkx}
\bibinfo{author}{\bibfnamefont{I.}~\bibnamefont{Adachi}} \bibnamefont{et~al.} (\bibinfo{collaboration}{ILC International Community}) (\bibinfo{year}{2022}), \eprint{2203.07622}.

\bibitem[{\citenamefont{Celiberto et~al.}(2021{\natexlab{b}})}]{Celiberto:2021dzy}
\bibinfo{author}{\bibfnamefont{F.~G.} \bibnamefont{Celiberto}} \bibnamefont{et~al.}, \bibinfo{journal}{Eur. Phys. J. C} \textbf{\bibinfo{volume}{81}}, \bibinfo{pages}{780} (\bibinfo{year}{2021}{\natexlab{b}}), \eprint{2105.06432}.

\bibitem[{\citenamefont{Celiberto et~al.}(2021{\natexlab{c}})}]{Celiberto:2021fdp}
\bibinfo{author}{\bibfnamefont{F.~G.} \bibnamefont{Celiberto}} \bibnamefont{et~al.}, \bibinfo{journal}{Phys. Rev. D} \textbf{\bibinfo{volume}{104}}, \bibinfo{pages}{114007} (\bibinfo{year}{2021}{\natexlab{c}}), \eprint{2109.11875}.

\bibitem[{\citenamefont{Celiberto et~al.}(2022{\natexlab{b}})}]{Celiberto:2022zdg}
\bibinfo{author}{\bibfnamefont{F.~G.} \bibnamefont{Celiberto}} \bibnamefont{et~al.}, \bibinfo{journal}{Phys. Rev. D} \textbf{\bibinfo{volume}{105}}, \bibinfo{pages}{114056} (\bibinfo{year}{2022}{\natexlab{b}}), \eprint{2205.13429}.

\bibitem[{\citenamefont{Celiberto}(2022{\natexlab{b}})}]{Celiberto:2022keu}
\bibinfo{author}{\bibfnamefont{F.~G.} \bibnamefont{Celiberto}}, \bibinfo{journal}{Phys. Lett. B} \textbf{\bibinfo{volume}{835}}, \bibinfo{pages}{137554} (\bibinfo{year}{2022}{\natexlab{b}}), \eprint{2206.09413}.

\bibitem[{\citenamefont{Anchordoqui et~al.}(2022)}]{Anchordoqui:2021ghd}
\bibinfo{author}{\bibfnamefont{L.~A.} \bibnamefont{Anchordoqui}} \bibnamefont{et~al.}, \bibinfo{journal}{Phys. Rept.} \textbf{\bibinfo{volume}{968}}, \bibinfo{pages}{1} (\bibinfo{year}{2022}), \eprint{2109.10905}.

\bibitem[{\citenamefont{Feng et~al.}(2023)}]{Feng:2022inv}
\bibinfo{author}{\bibfnamefont{J.~L.} \bibnamefont{Feng}} \bibnamefont{et~al.}, \bibinfo{journal}{J. Phys. G} \textbf{\bibinfo{volume}{50}}, \bibinfo{pages}{030501} (\bibinfo{year}{2023}), \eprint{2203.05090}.

\bibitem[{\citenamefont{Boussarie et~al.}(2018)}]{Boussarie:2017oae}
\bibinfo{author}{\bibfnamefont{R.}~\bibnamefont{Boussarie}} \bibnamefont{et~al.}, \bibinfo{journal}{Phys. Rev. D} \textbf{\bibinfo{volume}{97}}, \bibinfo{pages}{014008} (\bibinfo{year}{2018}), \eprint{1709.01380}.

\bibitem[{\citenamefont{Chapon et~al.}(2022)}]{Chapon:2020heu}
\bibinfo{author}{\bibfnamefont{E.}~\bibnamefont{Chapon}} \bibnamefont{et~al.}, \bibinfo{journal}{Prog. Part. Nucl. Phys.} \textbf{\bibinfo{volume}{122}}, \bibinfo{pages}{103906} (\bibinfo{year}{2022}), \eprint{2012.14161}.

\bibitem[{\citenamefont{Celiberto  Fucilla}(2022)}]{Celiberto:2022dyf}
\bibinfo{author}{\bibfnamefont{F.~G.} \bibnamefont{Celiberto}}  \bibinfo{author}{\bibfnamefont{M.}~\bibnamefont{Fucilla}}, \bibinfo{journal}{Eur. Phys. J. C} \textbf{\bibinfo{volume}{82}}, \bibinfo{pages}{929} (\bibinfo{year}{2022}), \eprint{2202.12227}.

\bibitem[{\citenamefont{Celiberto}(2023{\natexlab{b}})}]{Celiberto:2023fzz}
\bibinfo{author}{\bibfnamefont{F.~G.} \bibnamefont{Celiberto}}, \bibinfo{journal}{Universe} \textbf{\bibinfo{volume}{9}}, \bibinfo{pages}{324} (\bibinfo{year}{2023}{\natexlab{b}}), \eprint{2305.14295}.

\bibitem[{\citenamefont{Stebel  Watanabe}(2021)}]{Stebel:2021bbn}
\bibinfo{author}{\bibfnamefont{T.}~\bibnamefont{Stebel}}  \bibinfo{author}{\bibfnamefont{K.}~\bibnamefont{Watanabe}}, \bibinfo{journal}{Phys. Rev. D} \textbf{\bibinfo{volume}{104}}, \bibinfo{pages}{034004} (\bibinfo{year}{2021}), \eprint{2103.01724}.

\bibitem[{\citenamefont{Celiberto  Papa}(2023{\natexlab{a}})}]{Celiberto:2023rzw}
\bibinfo{author}{\bibfnamefont{F.~G.} \bibnamefont{Celiberto}}  \bibinfo{author}{\bibfnamefont{A.}~\bibnamefont{Papa}} (\bibinfo{year}{2023}{\natexlab{a}}), \eprint{2308.00809}.

\bibitem[{\citenamefont{Boughezal et~al.}(2013)}]{Boughezal:2013uia}
\bibinfo{author}{\bibfnamefont{R.}~\bibnamefont{Boughezal}} \bibnamefont{et~al.}, \bibinfo{journal}{JHEP} \textbf{\bibinfo{volume}{06}}, \bibinfo{pages}{072} (\bibinfo{year}{2013}), \eprint{1302.6216}.

\bibitem[{\citenamefont{Chen et~al.}(2015)}]{Chen:2014gva}
\bibinfo{author}{\bibfnamefont{X.}~\bibnamefont{Chen}} \bibnamefont{et~al.}, \bibinfo{journal}{Phys. Lett. B} \textbf{\bibinfo{volume}{740}}, \bibinfo{pages}{147} (\bibinfo{year}{2015}), \eprint{1408.5325}.

\bibitem[{\citenamefont{Boughezal et~al.}(2015)}]{Boughezal:2015dra}
\bibinfo{author}{\bibfnamefont{R.}~\bibnamefont{Boughezal}} \bibnamefont{et~al.}, \bibinfo{journal}{Phys. Rev. Lett.} \textbf{\bibinfo{volume}{115}}, \bibinfo{pages}{082003} (\bibinfo{year}{2015}), \eprint{1504.07922}.

\bibitem[{\citenamefont{Monni et~al.}(2020)\citenamefont{Monni, Rottoli,  Torrielli}}]{Monni:2019yyr}
\bibinfo{author}{\bibfnamefont{P.~F.} \bibnamefont{Monni}}, \bibinfo{author}{\bibfnamefont{L.}~\bibnamefont{Rottoli}},  \bibinfo{author}{\bibfnamefont{P.}~\bibnamefont{Torrielli}}, \bibinfo{journal}{Phys. Rev. Lett.} \textbf{\bibinfo{volume}{124}}, \bibinfo{pages}{252001} (\bibinfo{year}{2020}), \eprint{1909.04704}.

\bibitem[{\citenamefont{Celiberto  Papa}(2023{\natexlab{b}})}]{Celiberto:2023rtu}
\bibinfo{author}{\bibfnamefont{F.~G.} \bibnamefont{Celiberto}}  \bibinfo{author}{\bibfnamefont{A.}~\bibnamefont{Papa}} (\bibinfo{year}{2023}{\natexlab{b}}), \eprint{2305.00962}.

\bibitem[{\citenamefont{Del~Duca  Schmidt}(1994)}]{DelDuca:1993ga}
\bibinfo{author}{\bibfnamefont{V.}~\bibnamefont{Del~Duca}}  \bibinfo{author}{\bibfnamefont{C.~R.} \bibnamefont{Schmidt}}, \bibinfo{journal}{Phys. Rev. D} \textbf{\bibinfo{volume}{49}}, \bibinfo{pages}{177} (\bibinfo{year}{1994}), \eprint{hep-ph/9305346}.

\bibitem[{\citenamefont{Celiberto et~al.}(2023)}]{Celiberto:2023uuk_article}
\bibinfo{author}{\bibfnamefont{F.~G.} \bibnamefont{Celiberto}} \bibnamefont{et~al.} (\bibinfo{year}{2023}), \eprint{2305.05052}.

\bibitem[{\citenamefont{Celiberto et~al.}(2022{\natexlab{c}})}]{Celiberto:2022fgx}
\bibinfo{author}{\bibfnamefont{F.~G.} \bibnamefont{Celiberto}} \bibnamefont{et~al.}, \bibinfo{journal}{JHEP} \textbf{\bibinfo{volume}{08}}, \bibinfo{pages}{092} (\bibinfo{year}{2022}{\natexlab{c}}), \eprint{2205.02681}.

\bibitem[{\citenamefont{Fucilla}(2023{\natexlab{a}})}]{Fucilla:2023pma}
\bibinfo{author}{\bibfnamefont{M.}~\bibnamefont{Fucilla}}, Ph.D. thesis (\bibinfo{year}{2023}{\natexlab{a}}), \eprint{2308.03393}.

\bibitem[{\citenamefont{Fucilla}(2023{\natexlab{b}})}]{Fucilla:2022whr}
\bibinfo{author}{\bibfnamefont{M.}~\bibnamefont{Fucilla}}, \bibinfo{journal}{Acta Phys. Polon. Supp.} \textbf{\bibinfo{volume}{16}}, \bibinfo{pages}{44} (\bibinfo{year}{2023}{\natexlab{b}}), \eprint{2212.01794}.

\bibitem[{\citenamefont{Hentschinski et~al.}(2021)}]{Hentschinski:2020tbi}
\bibinfo{author}{\bibfnamefont{M.}~\bibnamefont{Hentschinski}} \bibnamefont{et~al.}, \bibinfo{journal}{Eur. Phys. J. C} \textbf{\bibinfo{volume}{81}}, \bibinfo{pages}{112} (\bibinfo{year}{2021}), \eprint{2011.03193}.

\bibitem[{\citenamefont{Celiberto}(2021{\natexlab{b}})}]{Celiberto:2020wpk}
\bibinfo{author}{\bibfnamefont{F.~G.} \bibnamefont{Celiberto}}, \bibinfo{journal}{Eur. Phys. J. C} \textbf{\bibinfo{volume}{81}}, \bibinfo{pages}{691} (\bibinfo{year}{2021}{\natexlab{b}}), \eprint{2008.07378}.

\bibitem[{\citenamefont{Celiberto}(2022{\natexlab{c}})}]{Celiberto:2022rfj}
\bibinfo{author}{\bibfnamefont{F.~G.} \bibnamefont{Celiberto}}, \bibinfo{journal}{Phys. Rev. D} \textbf{\bibinfo{volume}{105}}, \bibinfo{pages}{114008} (\bibinfo{year}{2022}{\natexlab{c}}), \eprint{2204.06497}.

\bibitem[{\citenamefont{Nason}(2004)}]{Nason:2004rx}
\bibinfo{author}{\bibfnamefont{P.}~\bibnamefont{Nason}}, \bibinfo{journal}{JHEP} \textbf{\bibinfo{volume}{11}}, \bibinfo{pages}{040} (\bibinfo{year}{2004}), \eprint{hep-ph/0409146}.

\bibitem[{\citenamefont{Campbell et~al.}(2012)}]{Campbell:2012am}
\bibinfo{author}{\bibfnamefont{J.~M.} \bibnamefont{Campbell}} \bibnamefont{et~al.}, \bibinfo{journal}{JHEP} \textbf{\bibinfo{volume}{07}}, \bibinfo{pages}{092} (\bibinfo{year}{2012}), \eprint{1202.5475}.

\bibitem[{\citenamefont{Hamilton et~al.}(2013)}]{Hamilton:2012rf}
\bibinfo{author}{\bibfnamefont{K.}~\bibnamefont{Hamilton}} \bibnamefont{et~al.}, \bibinfo{journal}{JHEP} \textbf{\bibinfo{volume}{05}}, \bibinfo{pages}{082} (\bibinfo{year}{2013}), \eprint{1212.4504}.

\bibitem[{\citenamefont{Banfi et~al.}(2023)}]{Banfi:2023mhz}
\bibinfo{author}{\bibfnamefont{A.}~\bibnamefont{Banfi}} \bibnamefont{et~al.} (\bibinfo{year}{2023}), \eprint{2309.02127}.

\bibitem[{\citenamefont{Bagnaschi et~al.}(2023)\citenamefont{Bagnaschi, Degrassi,  Gr\"ober}}]{Bagnaschi:2023rbx}
\bibinfo{author}{\bibfnamefont{E.}~\bibnamefont{Bagnaschi}}, \bibinfo{author}{\bibfnamefont{G.}~\bibnamefont{Degrassi}},  \bibinfo{author}{\bibfnamefont{R.}~\bibnamefont{Gr\"ober}} (\bibinfo{year}{2023}), \eprint{2309.10525}.

\bibitem[{\citenamefont{Alioli et~al.}(2023{\natexlab{a}})}]{Alioli:2022dkj}
\bibinfo{author}{\bibfnamefont{S.}~\bibnamefont{Alioli}} \bibnamefont{et~al.}, \bibinfo{journal}{JHEP} \textbf{\bibinfo{volume}{06}}, \bibinfo{pages}{205} (\bibinfo{year}{2023}{\natexlab{a}}), \eprint{2212.10489}.

\bibitem[{\citenamefont{Alioli et~al.}(2023{\natexlab{b}})}]{Alioli:2023har}
\bibinfo{author}{\bibfnamefont{S.}~\bibnamefont{Alioli}} \bibnamefont{et~al.}, \bibinfo{journal}{JHEP} \textbf{\bibinfo{volume}{05}}, \bibinfo{pages}{128} (\bibinfo{year}{2023}{\natexlab{b}}), \eprint{2301.11875}.

\bibitem[{\citenamefont{Buckley et~al.}(2021)}]{Buckley:2021gfw}
\bibinfo{author}{\bibfnamefont{A.}~\bibnamefont{Buckley}} \bibnamefont{et~al.}, \bibinfo{journal}{JHEP} \textbf{\bibinfo{volume}{11}}, \bibinfo{pages}{108} (\bibinfo{year}{2021}), \eprint{2105.11399}.

\bibitem[{\citenamefont{van Beekveld et~al.}(2022{\natexlab{a}})}]{vanBeekveld:2022zhl}
\bibinfo{author}{\bibfnamefont{M.}~\bibnamefont{van Beekveld}} \bibnamefont{et~al.}, \bibinfo{journal}{JHEP} \textbf{\bibinfo{volume}{11}}, \bibinfo{pages}{019} (\bibinfo{year}{2022}{\natexlab{a}}), \eprint{2205.02237}.

\bibitem[{\citenamefont{van Beekveld et~al.}(2022{\natexlab{b}})}]{vanBeekveld:2022ukn}
\bibinfo{author}{\bibfnamefont{M.}~\bibnamefont{van Beekveld}} \bibnamefont{et~al.}, \bibinfo{journal}{JHEP} \textbf{\bibinfo{volume}{11}}, \bibinfo{pages}{020} (\bibinfo{year}{2022}{\natexlab{b}}), \eprint{2207.09467}.

\bibitem[{\citenamefont{Ferrario~Ravasio et~al.}(2023)}]{FerrarioRavasio:2023kyg}
\bibinfo{author}{\bibfnamefont{S.}~\bibnamefont{Ferrario~Ravasio}} \bibnamefont{et~al.} (\bibinfo{year}{2023}), \eprint{2307.11142}.

\bibitem[{\citenamefont{Ball et~al.}(2021)}]{NNPDF:2021uiq}
\bibinfo{author}{\bibfnamefont{R.~D.} \bibnamefont{Ball}} \bibnamefont{et~al.} (\bibinfo{collaboration}{NNPDF}), \bibinfo{journal}{Eur. Phys. J. C} \textbf{\bibinfo{volume}{81}}, \bibinfo{pages}{958} (\bibinfo{year}{2021}), \eprint{2109.02671}.

\bibitem[{\citenamefont{Ball et~al.}(2022)}]{NNPDF:2021njg}
\bibinfo{author}{\bibfnamefont{R.~D.} \bibnamefont{Ball}} \bibnamefont{et~al.} (\bibinfo{collaboration}{NNPDF}), \bibinfo{journal}{Eur. Phys. J. C} \textbf{\bibinfo{volume}{82}}, \bibinfo{pages}{428} (\bibinfo{year}{2022}), \eprint{2109.02653}.

\bibitem[{\citenamefont{Ball et~al.}(2018)}]{Ball:2017otu}
\bibinfo{author}{\bibfnamefont{R.~D.} \bibnamefont{Ball}} \bibnamefont{et~al.}, \bibinfo{journal}{Eur. Phys. J.} \textbf{\bibinfo{volume}{C78}}, \bibinfo{pages}{321} (\bibinfo{year}{2018}), \eprint{1710.05935}.

\bibitem[{\citenamefont{Bonvini et~al.}(2015)}]{Bonvini:2015ira}
\bibinfo{author}{\bibfnamefont{M.}~\bibnamefont{Bonvini}} \bibnamefont{et~al.}, \bibinfo{journal}{JHEP} \textbf{\bibinfo{volume}{09}}, \bibinfo{pages}{191} (\bibinfo{year}{2015}), \eprint{1507.01006}.

\bibitem[{\citenamefont{Sterman}(1987)}]{Sterman:1986aj}
\bibinfo{author}{\bibfnamefont{G.~F.} \bibnamefont{Sterman}}, \bibinfo{journal}{Nucl. Phys. B} \textbf{\bibinfo{volume}{281}}, \bibinfo{pages}{310} (\bibinfo{year}{1987}).

\bibitem[{\citenamefont{Catani  Trentadue}(1989)}]{Catani:1989ne}
\bibinfo{author}{\bibfnamefont{S.}~\bibnamefont{Catani}}  \bibinfo{author}{\bibfnamefont{L.}~\bibnamefont{Trentadue}}, \bibinfo{journal}{Nucl. Phys. B} \textbf{\bibinfo{volume}{327}}, \bibinfo{pages}{323} (\bibinfo{year}{1989}).

\bibitem[{\citenamefont{Catani et~al.}(1996)}]{Catani:1996yz}
\bibinfo{author}{\bibfnamefont{S.}~\bibnamefont{Catani}} \bibnamefont{et~al.}, \bibinfo{journal}{Nucl. Phys. B} \textbf{\bibinfo{volume}{478}}, \bibinfo{pages}{273} (\bibinfo{year}{1996}), \eprint{hep-ph/9604351}.

\bibitem[{\citenamefont{Bonciani et~al.}(2003)}]{Bonciani:2003nt}
\bibinfo{author}{\bibfnamefont{R.}~\bibnamefont{Bonciani}} \bibnamefont{et~al.}, \bibinfo{journal}{Phys. Lett. B} \textbf{\bibinfo{volume}{575}}, \bibinfo{pages}{268} (\bibinfo{year}{2003}), \eprint{hep-ph/0307035}.

\bibitem[{\citenamefont{de~Florian  Grazzini}(2012)}]{deFlorian:2012yg}
\bibinfo{author}{\bibfnamefont{D.}~\bibnamefont{de~Florian}}  \bibinfo{author}{\bibfnamefont{M.}~\bibnamefont{Grazzini}}, \bibinfo{journal}{Phys. Lett. B} \textbf{\bibinfo{volume}{718}}, \bibinfo{pages}{117} (\bibinfo{year}{2012}), \eprint{1206.4133}.

\bibitem[{\citenamefont{Forte et~al.}(2021)\citenamefont{Forte, Ridolfi,  Rota}}]{Forte:2021wxe}
\bibinfo{author}{\bibfnamefont{S.}~\bibnamefont{Forte}}, \bibinfo{author}{\bibfnamefont{G.}~\bibnamefont{Ridolfi}},  \bibinfo{author}{\bibfnamefont{S.}~\bibnamefont{Rota}}, \bibinfo{journal}{JHEP} \textbf{\bibinfo{volume}{08}}, \bibinfo{pages}{110} (\bibinfo{year}{2021}), \eprint{2106.11321}.

\bibitem[{\citenamefont{Mukherjee  Vogelsang}(2006)}]{Mukherjee:2006uu}
\bibinfo{author}{\bibfnamefont{A.}~\bibnamefont{Mukherjee}}  \bibinfo{author}{\bibfnamefont{W.}~\bibnamefont{Vogelsang}}, \bibinfo{journal}{Phys. Rev. D} \textbf{\bibinfo{volume}{73}}, \bibinfo{pages}{074005} (\bibinfo{year}{2006}), \eprint{hep-ph/0601162}.

\bibitem[{\citenamefont{Becher  Neubert}(2006)}]{Becher:2006nr}
\bibinfo{author}{\bibfnamefont{T.}~\bibnamefont{Becher}}  \bibinfo{author}{\bibfnamefont{M.}~\bibnamefont{Neubert}}, \bibinfo{journal}{Phys. Rev. Lett.} \textbf{\bibinfo{volume}{97}}, \bibinfo{pages}{082001} (\bibinfo{year}{2006}), \eprint{hep-ph/0605050}.

\bibitem[{\citenamefont{Becher et~al.}(2008)\citenamefont{Becher, Neubert,  Xu}}]{Becher:2007ty}
\bibinfo{author}{\bibfnamefont{T.}~\bibnamefont{Becher}}, \bibinfo{author}{\bibfnamefont{M.}~\bibnamefont{Neubert}},  \bibinfo{author}{\bibfnamefont{G.}~\bibnamefont{Xu}}, \bibinfo{journal}{JHEP} \textbf{\bibinfo{volume}{07}}, \bibinfo{pages}{030} (\bibinfo{year}{2008}), \eprint{0710.0680}.

\bibitem[{\citenamefont{Bonvini et~al.}(2011)\citenamefont{Bonvini, Forte,  Ridolfi}}]{Bonvini:2010tp}
\bibinfo{author}{\bibfnamefont{M.}~\bibnamefont{Bonvini}}, \bibinfo{author}{\bibfnamefont{S.}~\bibnamefont{Forte}},  \bibinfo{author}{\bibfnamefont{G.}~\bibnamefont{Ridolfi}}, \bibinfo{journal}{Nucl. Phys. B} \textbf{\bibinfo{volume}{847}}, \bibinfo{pages}{93} (\bibinfo{year}{2011}), \eprint{1009.5691}.

\bibitem[{\citenamefont{Ahmed et~al.}(2015)}]{Ahmed:2014era}
\bibinfo{author}{\bibfnamefont{T.}~\bibnamefont{Ahmed}} \bibnamefont{et~al.}, \bibinfo{journal}{JHEP} \textbf{\bibinfo{volume}{02}}, \bibinfo{pages}{131} (\bibinfo{year}{2015}), \eprint{1411.5301}.

\bibitem[{\citenamefont{Banerjee et~al.}(2018)}]{Banerjee:2018vvb}
\bibinfo{author}{\bibfnamefont{P.}~\bibnamefont{Banerjee}} \bibnamefont{et~al.}, \bibinfo{journal}{Phys. Rev. D} \textbf{\bibinfo{volume}{98}}, \bibinfo{pages}{054018} (\bibinfo{year}{2018}), \eprint{1805.01186}.

\bibitem[{\citenamefont{Ajjath et~al.}(2022)}]{Ajjath:2021bbm}
\bibinfo{author}{\bibfnamefont{A.~H.} \bibnamefont{Ajjath}} \bibnamefont{et~al.}, \bibinfo{journal}{Eur. Phys. J. C} \textbf{\bibinfo{volume}{82}}, \bibinfo{pages}{774} (\bibinfo{year}{2022}), \eprint{2109.12657}.

\bibitem[{\citenamefont{Duhr et~al.}(2022)\citenamefont{Duhr, Mistlberger,  Vita}}]{Duhr:2022cob}
\bibinfo{author}{\bibfnamefont{C.}~\bibnamefont{Duhr}}, \bibinfo{author}{\bibfnamefont{B.}~\bibnamefont{Mistlberger}},  \bibinfo{author}{\bibfnamefont{G.}~\bibnamefont{Vita}}, \bibinfo{journal}{JHEP} \textbf{\bibinfo{volume}{09}}, \bibinfo{pages}{155} (\bibinfo{year}{2022}), \eprint{2205.04493}.

\bibitem[{\citenamefont{Jones et~al.}(2018)\citenamefont{Jones, Kerner,  Luisoni}}]{Jones:2018hbb}
\bibinfo{author}{\bibfnamefont{S.~P.} \bibnamefont{Jones}}, \bibinfo{author}{\bibfnamefont{M.}~\bibnamefont{Kerner}},  \bibinfo{author}{\bibfnamefont{G.}~\bibnamefont{Luisoni}}, \bibinfo{journal}{Phys. Rev. Lett.} \textbf{\bibinfo{volume}{120}}, \bibinfo{pages}{162001} (\bibinfo{year}{2018}), \eprint{1802.00349}.

\bibitem[{\citenamefont{Bonciani et~al.}(2023)}]{Bonciani:2022jmb}
\bibinfo{author}{\bibfnamefont{R.}~\bibnamefont{Bonciani}} \bibnamefont{et~al.}, \bibinfo{journal}{Phys. Lett. B} \textbf{\bibinfo{volume}{843}}, \bibinfo{pages}{137995} (\bibinfo{year}{2023}), \eprint{2206.10490}.

\bibitem[{\citenamefont{Andersen et~al.}(2023{\natexlab{a}})}]{Andersen:2022zte}
\bibinfo{author}{\bibfnamefont{J.~R.} \bibnamefont{Andersen}} \bibnamefont{et~al.}, \bibinfo{journal}{JHEP} \textbf{\bibinfo{volume}{03}}, \bibinfo{pages}{001} (\bibinfo{year}{2023}{\natexlab{a}}), \eprint{2210.10671}.

\bibitem[{\citenamefont{Andersen et~al.}(2023{\natexlab{b}})}]{Andersen:2023kuj}
\bibinfo{author}{\bibfnamefont{J.~R.} \bibnamefont{Andersen}} \bibnamefont{et~al.} (\bibinfo{year}{2023}{\natexlab{b}}), \eprint{2303.15778}.

\end{thebibliography}
\endgroup

\end{document}